\documentclass[a4paper,11pt]{article}
\usepackage{wrapfig2}
\usepackage{pos}
\usepackage{plex-serif,plex-sans,plex-mono}
\usepackage{amsmath,amsfonts,bbm,caption,float,graphicx,hyperref,mathrsfs,multicol}
\usepackage[british]{babel}
\usepackage[T1]{fontenc}
\DeclareMathOperator{\Tr}{Tr}
\DeclareMathOperator{\RE}{Re}
\usepackage{mdwlist,relsize,siunitx,subcaption}

\newcommand{\centring}{\centering}

\usepackage[inkscapelatex=false]{svg}
\svgsetup{clean=false}
\newcommand{\cprev}{\cite{HandsEtAl2006,HandsEtAl2010,BozEtAl2013EPJA,CotterEtAl2013PhysRev,CottThesis,TamerThesis,
							BozEtAl2020}}

\newcommand{\ccott}{\cite{BozEtAl2013EPJA,CotterEtAl2013PhysRev,CottThesis}}

\newcommand{\rjapan}{\cite{Iida:2024irv,Itou:2023pcl,Iida:2022hyy}}
\newcommand{\ctwo}{\cite{HandsEtAl2006,HandsEtAl2010,BozEtAl2013EPJA,CotterEtAl2013PhysRev,CottThesis,
							TamerThesis,BozEtAl2020,Iida:2024irv,Itou:2023pcl,Iida:2022hyy,Begun:2022bxj}}

\graphicspath{{./Images/}}
\svgpath{{./Plots/}}
\usepackage[automake,acronym,symbols,nomain,nogroupskip,nonumberlist,sort=use,record,shortcuts=ac]{glossaries-extra}
\GlsXtrLoadResources[
	src={Glossary},
	sort={en-GB},
	type=acronym,
	save-locations=false
]

\begin{document}
\title{Dense \ac{qc2d}. What's up with that?!?}
\author[a]{Simon Hands}
\author[b]{Seyong Kim}
\author*[c]{Dale Lawlor}
\author[c,d]{Andrew Lee-Mitchell}
\author[c,e]{Jon-Ivar Skullerud}
\affiliation[a]{Department of Mathematical Sciences, University of Liverpool,Liverpool,L69 3BX,UK}
\affiliation[b]{Department of Physics, Sejong University,Seoul 143--147,KR}
\affiliation[c]{Department of Theoretical Physics, National University of Ireland Maynooth,Maynooth,Kildare,IE}
\affiliation[d]{Department of Electronic Engineering, National University of Ireland Maynooth,Maynooth,Kildare,IE}
\affiliation[e]{Hamilton Institute, National University of Ireland Maynooth,Maynooth,Kildare,IE}

\abstract{
We present recent updates and results from \Acf{qc2d} simulations at non-zero baryon density,
including progress towards determining the speed of sound.
}
\emailAdd{simon.hands@liverpool.ac.uk}
\emailAdd{skim@sejong.ac.kr}
\emailAdd{dalel487@thphys.nuim.ie}
\emailAdd{andrew.leemitchell.2021@mumail.ie}
\emailAdd{jonivar@thphys.nuim.ie}

\FullConference{The 41st International Symposium on Lattice Field Theory (LATTICE2024)\\
28 July - 3 August 2024\\
Liverpool, UK\\}

\maketitle
\section{Introduction}\label{Sec:Intro}
	Lattice simulations of \Acf{qcd} at finite density are complicated by the complex fermion action inhibiting
	importance sampling. Fortunately, there are several \ac{qcd}-like theories such as \ac{qc2d} \ctwo, isospin \ac{qcd}
	\cite{Brandt:2022hwy,Brandt:2017oyy,Brandt:2018bwq} or imaginary chemical potential studies which give insights into
	the behaviour of real \ac{qcd}. Other lattice approaches are summarised nicely in \cite{Nagata:2021ugx}. This work
	focuses on \Acf{qc2d} where we replace the $SU(3)$ gauge theory with an $SU(2)$ one. This does not exhibit a complex
	action problem for an even number of quark flavours\footnote{See \cite{Hands:2000ei} for a proof}.

	Once one has settled on how best to simulate at finite density, the next thing to decide on is the best probe to
	study the finite density r\'egime. An overview of non-lattice studies of dense \ac{qcd} matter can be found in
	\cite{Koehn:2024set}, but one popular method of exploring the \ac{qcd} \glsxtrfull{eos} is to look at how the
	speed of sound
	\begin{equation}\label{Eq:C_s}
			C_s^2=\frac{\partial P}{\partial\varepsilon}
	\end{equation}
	behaves as density increases. Whilst it has been predicted from perturbation that $C_s^2$ approaches the conformal
	limit $\frac{1}{3}$ from below recent non-perturbative results \cite{Iida:2024irv,Iida:2022hyy,Brandt:2022hwy},
	perturbative studies \cite{Koehn:2024set} and astronomical observations
	\cite{Riley:2019yda,Miller:2019cac,LIGOScientific:2018cki,LIGOScientific:2017vwq} appear to indicate otherwise. At
	the other end of the scale recent work in relativistic hydrodynamics suggest that there is an upper bound on $C_s^2$
	from the shear viscosity \cite{Hippert:2024hum}.

	This work has two objectives. To calculate the \ac{qc2d} speed of sound on a finer lattice than previously
	studied \rjapan{} and to consider a larger number of diquark sources $aj$ at smaller values to get better control over
	the diquark source extrapolation.
\section{Scale Setting and Beta Functions}\label{Sec:Scale}
	Calculating $C_s^2$ requires evaluating the pressure $P$ and energy density which shall be discussed further in section
	\ref{Sec:Sim}. In this work we determine the energy density via
	\begin{align}\label{Eq:EnDen}
		\varepsilon=T_{\mu\mu}+3P\\
\label{Eq:TraceAnom}
	 	T_{\mu\mu}=T^g_{\mu\mu}+T^q_{\mu\mu}
	\end{align}
	Both $T^g_{\mu\mu}$ and $T^q_{\mu\mu}$ require renormalisation by the beta functions which shall be discussed in
	subsections \ref{Sub:Mass} and \ref{Sub:Space}.

	The beta functions in previous works \ccott{} were derived using the Karsch coefficients, which requires anisotropic
	lattices (see the first three rows of table \ref{Tab:Const_phys}). For this work we instead determined
	$\beta$--$\kappa$ pairs that lie on the line of constant physics $\frac{m_\pi}{m_\rho}\sim 0.81$. This requires
	multiple rounds of scale setting as seen in table \ref{Tab:Const_phys}.
\subsection{Scale Setting: Mass Tuning}\label{Sub:Mass}
	Firstly, for a chosen $\beta$ we make an educated guess for what $\kappa$ will give
	the correct mass ratio. We then produce $\sim 500$ trajectories (saving every $5$th configuration). As a non-physical
	theory there are no physical quantities that we can compare with to conduct scale setting. Instead we calculate the
	\emph{pion} (pseudoscalar) and \emph{rho} (vector) correlators, fit to

	\begin{equation}
		C(\tau)=C_0\cosh\left(M\left(\tau-\frac{N_\tau}{2}\right)\right)
	\end{equation}
	and extracting the mass in lattice units as $am=|M|$. If after taking the pion and rho mass ratio we lie on the
	desired line of constant physics we can continue producing more configurations to improve statistics and prepare to
	evaluate the lattice spacing.
\subsection{Scale Setting: Lattice Spacing}\label{Sub:Space}
	We determine the lattice spacing $a$ from the Cornell form of the \Acf{sqp}
	\begin{equation}\label{Eq:SQP}
		aV(r)=-\frac{N_c^2-1}{N_c}\frac{\alpha_s}{r}+a^2\sigma r+aV_0
	\end{equation}
	where $\alpha_s$ denotes the \ac{qcd} running coupling and $\sigma=(\qty{440}{\MeV})^2$ the string tension.
	Previous works by our group have	extracted the \ac{sqp} by evaluating and fitting to Wilson loops. However, 
	Wilson loops require path finding algorithms to evaluate and the cost scales poorly with the lattice volume. This work
	instead uses Wilson \emph{lines}
	\begin{equation}\label{Eq:Wilson Line}
		W(\vec{x}-\vec{y},t)=\Tr\left(U(\vec{x},t) U^\dagger(\vec{y},t)\right)
	\end{equation}
	and
	\begin{equation}\label{Eq:Temp Link}
		U(\vec{x},t)=\prod\limits_{j=t}^{t+l} U_4(\vec{x},j)
	\end{equation}
	where $0<l<\frac{N_t}{2}$ to extract the \ac{sqp}. Results in preparation indicate that the Wilson loop and Wilson
	line (fixed to Coulomb gauge) results agree within $2\sigma$, but with less computation needed.
\begin{figure}[h]
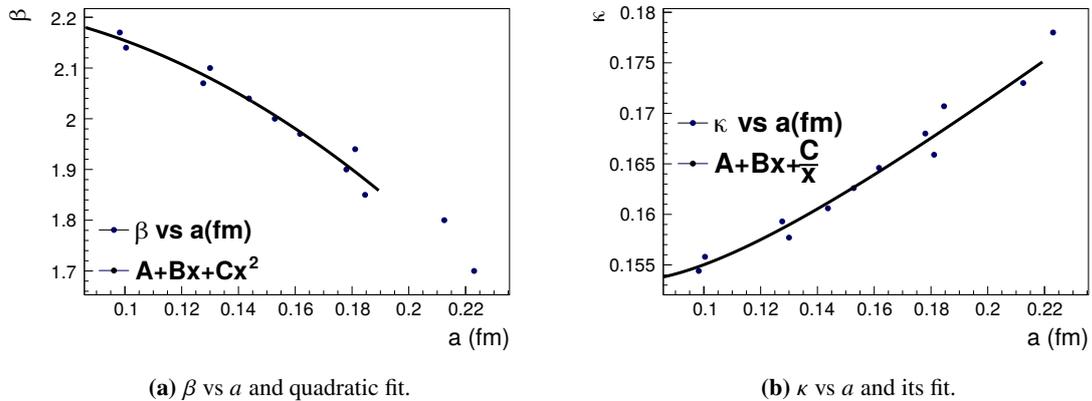

	\begin{subfigure}{0.5\linewidth}
		\centring
		\includesvg[width=1.0\linewidth,pretex=\relscale{0.8}]{Beta_vs_a}
		\subcaption{$\beta$ vs $a$ and quadratic fit.}
		\label{Fig:Beta_vs_a}
	\end{subfigure}
	\begin{subfigure}{0.5\linewidth}
		\centring
		\includesvg[width=1.0\linewidth,pretex=\relscale{0.8}]{kappa_vs_a}
		\subcaption{$\kappa$ vs $a$ and its fit.}
		\label{Fig:Kappa_vs_a}
	\end{subfigure}
	\caption{Fits to extract beta functions.}
	\label{Fig:Beta_Fits}
\end{figure}
\begin{table}
	\centring
		\begin{tabular}{|c|c|c|c|c|c|c|c|}
			\hline
			$\beta$ & $\kappa$ & $n_\text{cfg}$ & $a$ (\unit{\femto\metre}) & $\pm$ & $\frac{m_\pi}{m_\rho}$ & $\pm$ &$am_\pi$\\ \hline
			$\emph{1.7}$ & $\emph{0.1780}$ & & $\emph{0.223}$& $\emph{0.004}$ & $\emph{0.779}$ & $\emph{0.004}$ & $\emph{0.79}$	\\ \hline
			$\emph{1.9}$ & $\emph{0.1680}$ & & $\emph{0.187}$& $\emph{0.004}$ & $\emph{0.805}$ & $\emph{0.009}$ & $\emph{0.645}$	\\ \hline
			$\emph{2.1}$ & $\emph{0.1577}$ & & $0.130$ & $0.0004$ & $\emph{0.810}$ & $\emph{0.004}$ & $\emph{0.446}$	\\ \hline
			$1.80$ & $0.1730$ & $1019$  & $0.2125$ & $0.0019$ & $0.811$ & $0.008$ & $0.731$ \\ \hline
			$1.85$ & $0.1707$ & $277$  & $0.1846$ & $0.0015$ & $0.828$ & $0.021$ & $0.684$ \\ \hline
			$1.97$ & $0.1646$ & $175$  & $0.1617$ & $0.0008$ & $0.829$ & $0.027$ & $0.546$ \\ \hline
			$2.00$ & $0.1626$ & $268$  & $0.1528$ & $0.0006$ & $0.799$ & $0.022$ & $0.562$ \\ \hline
			$2.04$ & $0.1606$ & $293$  & $0.1437$ & $0.0004$ & $0.799$ & $0.018$ & $0.520$ \\ \hline
			$2.14$ & $0.1558$ & $106$  & $0.1004$ & $0.0007$ & $0.801$ & $0.045$ & $0.410$ \\ \hline
			$2.17$ & $0.1544$ & $100$  & $0.0741$ & $0.0016$ & $0.813$ & $0.009$ & $0.381$ \\ \hline
		\end{tabular}
		\caption{$\beta$ and $\kappa$ values on line of constant physics $\frac{m_\pi}{m_\rho}=0.81$. Italicised values are
		from earlier works.}
		\label{Tab:Const_phys}
\end{table}
\subsection{Remarks}\label{Sub:Remark}
	In table \ref{Tab:Beta_functions} we also present the beta functions corresponding to the coarse lattice used in
	\cprev. The values we obtained here are consistent with those obtained using the Karsch coefficients, which gives us
	confidence that the beta functions for the fine lattice used here are sane. This scale setting approach is different
	to that used in \rjapan{} where instead the spacing is tuned such that at $N_t=10$ at zero density the temperature
	$T=\qty{200}{\MeV}$.

\begin{table}
	\centring
		\begin{tabular}{|c|c|c|c|c|c|c|}
			\hline
			$\beta$ & $\kappa$ & $\frac{\partial\beta}{\partial a}$ &$\pm$ & $\frac{\partial \kappa}{\partial a}$ & $\pm$ &
			Remarks\\ \hline
			$1.9$ & $0.1680$ & $-2.71$ & $0.16$ & $0.197$ & $0.16$ & Karsch Coefficients in \cite{CottThesis} \\ \hline
			$1.9$ & $0.1680$ & $-2.86$ & $0.08$ & $0.195$ & $0.32$ & This work  \\ \hline
			$2.1$ & $0.1577$ & $-2.85$ & $0.10$ & $0.152$ & $0.32$ & This work \\ \hline
		\end{tabular}
		\caption{Beta functions as extracted from figure \ref{Fig:Beta_Fits}. The first two lines correspond to the coarse
		lattice used in \cprev. The last line are the values for the fine lattice used in this work.}
		\label{Tab:Beta_functions}
\end{table}
\section[Simulation and Results]{Simulation Details and Thermodynamic Observables}\label{Sec:Sim}
	We are using an unimproved Wilson fermion and gauge action for this simulation. At non-zero chemical potential the
	superfluid phase contains low-lying eigenmodes, so we introduce a \emph{diquark source} term $aj=0.01,0.015,0.02,0.03$
	to lift these modes making the simulation feasible and extrapolate to $j=0$. All the results shown here are from the
	\emph{fine} lattice where $\beta=2.1$ and $\kappa=0.1577$. This gives a spacing of $a=\qty{0.130}{\femto\metre}$. Our
	quark masses are tuned so that they lie on the line of constant physics $\frac{m_\pi}{m_\rho}=0.81$ discussed in
	section \ref{Sec:Scale}.

	We primarily choose chemical potential $a\mu_q$ at intervals of $0.050$, with increased resolution around the onset
	chemical potential $\mu_0=\frac{m_\pi}{2}=0.223$. Beyond $a\mu_q\sim0.75$ lattice artefacts render the results
	unreliable.  If we consider the $\frac{\mu_q}{m_\pi}$ axis we can explore higher chemical potentials than previous
	studies relative to the onset and critical chemical potentials due to the finer lattice spacing.
\subsection{Simulation Code}\label{SubSub:Code}
	The new gauge ensembles and scale setting data were produced over six weeks using the code in
	\cite{lawlor_2024_12910604}.  The major improvements over the original \cprev{} code are a mixed precision conjugate
	gradient, changing to the RANLUX \cite{Luscher:1993dy} generator, improved hybrid OpenMP/MPI support on CPU based
	machines and a CUDA port.  These ensembles were generated using the CUDA version of the code.

	The scale setting and analysis codes can also be found at \cite{skullerud_2024_14503846} alongside the data used in
	the speed of sound analysis at \cite{lawlor_2024_14201453}. The gauge configurations are not available on Zenodo but
	can be provided on request.
\subsection{Diquark Condensate}\label{Sub:Diquark}
	The diquark $\langle qq\rangle$ states are effectively the baryons of \ac{qc2d}. Unlike real
	\ac{qcd} these consist of only two quarks so are bosonic (the quarks themselves remain fermionic). The diquark
	condensate is strongly dependent on the diquark source so needs to be extrapolated to $aj=0$. We do this by fitting
	\begin{equation}\label{Eq:Diq_fit}
		\langle qq\rangle=C_0+C_1 j^{C_2}
	\end{equation}
	\Acf{chiPT} predicts that $C_2=\frac{1}{3}$ near onset. However at high densities this value is clearly violated.
	Above the onset chemical potential
	$\mu_0\sim\frac{m_\pi}{2}$ the diquark condensate takes on a non-zero value indicating the transition from a hadronic
	phase to a superfluid phase.
\begin{figure}
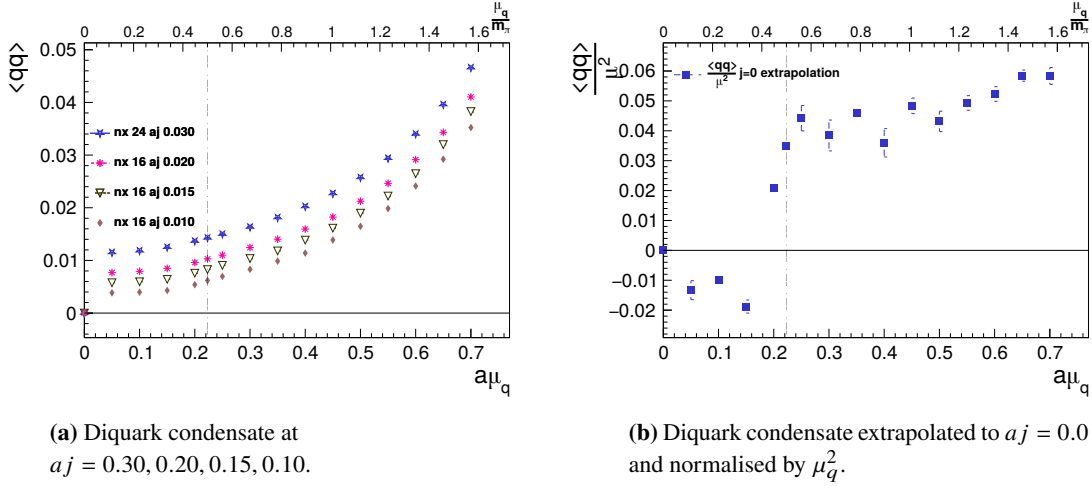

	\begin{subfigure}[t]{0.5\linewidth}
		\centring
		\captionsetup{width=.8\linewidth}
		\includesvg[width=1.0\linewidth,pretex=\relscale{0.8}]{qq.b210k1577t32}
		\caption{Diquark condensate at\\$aj=0.30, 0.20, 0.15, 0.10$.}
	\label{Fig:diq_no_extrap}
	\end{subfigure}
	\begin{subfigure}[t]{0.5\linewidth}
		\centring
		\captionsetup{width=.8\linewidth}
		\includesvg[width=1.0\linewidth,pretex=\relscale{0.8}]{qq_J_0_norm}
		\caption{Diquark condensate extrapolated to $aj=0.0$ and normalised by $\mu_q^2$.}
	\label{Fig:diq_norm}
	\end{subfigure}
	\caption{Diquark Condensate}
	\label{Fig:diq}
\end{figure}
\subsection{Quark Number Density and Pressure}\label{Sub:Pressure}
	The pressure is determined by integrating the quark number density with respect to $\mu_q$. Unlike $\langle
	qq\rangle$ it is only weakly dependent on the diquark source and is extrapolated to $aj=0$ using a linear fit. We
	also require the non-interacting lattice \Acf{sb} results to mitigate IR and UV artefacts in the pressure. The
	continuum and lattice \ac{sb} forms for non-interacting fermions can be found in \cite{BozEtAl2020} and
	\cite{HandsEtAl2006} respectively.
	\begin{align}
		\label{Eq:NII}
		n_q^\text{Lat}&=\frac{4N_fN_c}{N_s^3N_t}
		\sum\limits_k\frac{i\sin\tilde{k}_0\left[\sum\limits_i\cos{k_i}-\frac{1}{2\kappa}\right]}
		{\left[\frac{1}{2\kappa}-\sum\limits_\nu\cos\tilde{k}_\nu\right]^2+\sum\limits_\nu\sin^2\tilde{k}_\nu}
	\end{align}

	As was discussed in \cite{CotterEtAl2013PhysRev,BozEtAl2020} in order to evaluate $n_q^\text{Lat}$ one must consider
	a larger spatial volume than the one actually used. For this work, $4N_s$ was considered.
	We interpolate the quark number density using a cubic spline. We have two schemes to correct for lattice artefacts
	int the pressure as described in \cite{CotterEtAl2013PhysRev,BozEtAl2020}
	\begin{align}
		\label{Eq:PII}
		\text{Scheme II}:&\frac{P}{P_\text{SB}}(\mu_q)=\frac{1}{P_\text{Cont}(\mu_q)}\int\limits_0^{\mu_q}
		\frac{n_q^\text{Cont}}{n_q^\text{Lat}}\left(\mu'\right)n_q\left(\mu'\right)d\mu'
	\end{align}
	where the continuum \ac{sb} pressure is given by
	\begin{equation}
		P_\text{Cont}=\frac{N_fN_c}{12\pi^2}\left(\mu_q^4+2\pi^2\mu_q^2T^2+\frac{7\pi^4}{15}T^4\right)
	\end{equation}
\begin{figure}
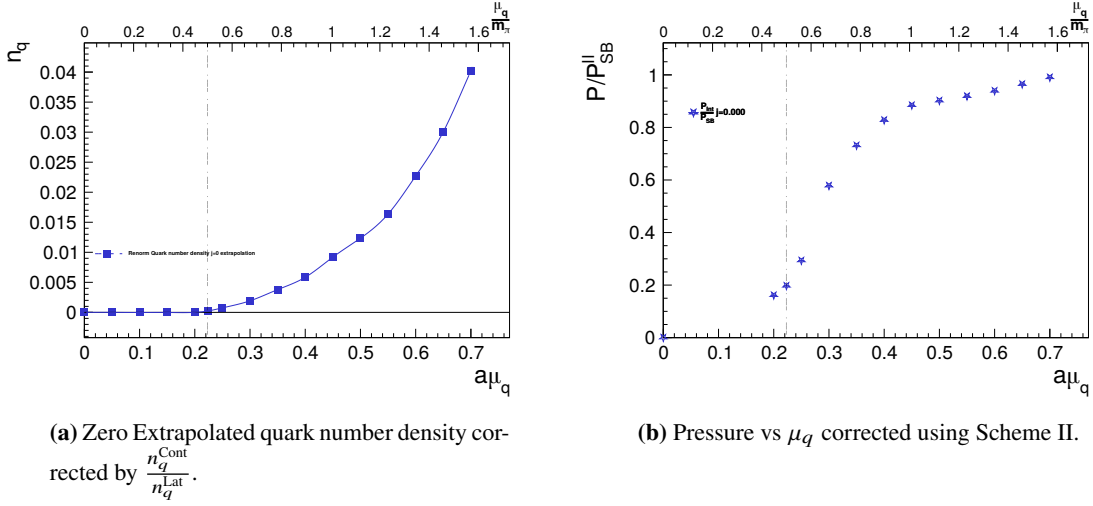

	\begin{subfigure}[t]{0.5\linewidth}
		\centring
		\captionsetup{width=.8\linewidth}
		\includesvg[width=1.0\linewidth,pretex=\relscale{0.8}]{n_q_II_J_0}
		\caption{Zero Extrapolated quark number density corrected by $\frac{n_q^\text{Cont}}{n_q^\text{Lat}}$.}
		\label{Fig:Norm n_q}
	\end{subfigure}
	\begin{subfigure}[t]{0.5\linewidth}
		\includesvg[width=1.0\linewidth,pretex=\relscale{0.8}]{PSB_v_mu_II.b210k1577j000s24t32}
		\captionsetup{width=.8\linewidth}
		\caption{Pressure vs $\mu_q$ corrected using Scheme II.}
		\label{Fig:Pressure Scheme II}
	\end{subfigure}
	\caption{Number density and pressure corrected by their \ac{sb} values.}
	\label{Fig:Norm Pressure}
\end{figure}
	Both the earlier results and this work indicate that Scheme II best approximates the \ac{sb} result at higher
	density. Thus we shall use it to evaluate $C_s^2$.
\subsection{Trace Anomaly}\label{Sub:Trace Anomaly}
	\begin{wrapfigure}[-2]{I}{0.5\textwidth}*
		\includesvg[width=1.0\linewidth,pretex=\relscale{0.8}]{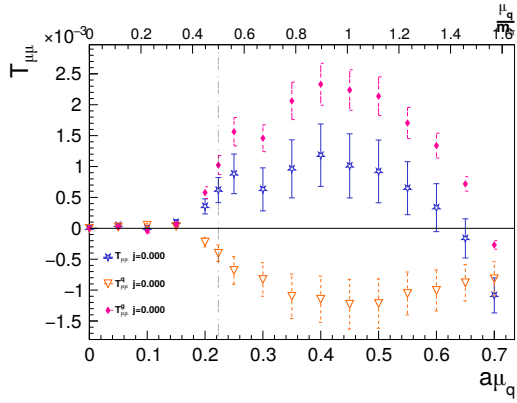}
		\caption{Trace Anomaly}
		\label{Fig:T_mumu}
	\end{wrapfigure}
	As mentioned in section \ref{Sec:Intro} we will derive $\varepsilon$ from the trace anomaly. From equation
	\eqref{Eq:TraceAnom} we see that the trace anomaly consists of a gluonic and a fermionic component given by
	\begin{align}\label{Eq:Trace Comp}
		T_{\mu\mu}^g&=-\frac{3a}{N_c}\frac{\partial\beta}{\partial a}\RE\left(\Tr{U_{ij}}+\Tr{U_{i0}}\right)\\
		T_{\mu\mu}^q&=-a\frac{\partial\kappa}{\partial a}\kappa^{-1}\left(4N_fN_c-\langle\bar{\psi}\psi\rangle\right)
	\end{align}

\begin{figure}[h]
	\begin{subfigure}[t]{0.5\linewidth}
		\centring
		\captionsetup{width=.8\linewidth}
		\includesvg[width=1.0\linewidth,pretex=\relscale{0.8}]{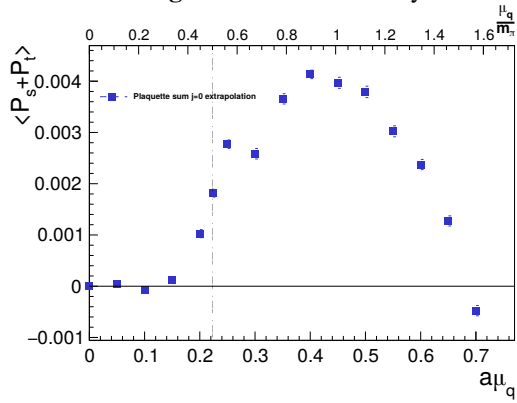}
		\caption{Subtracted and extrapolated plaquette sum.}
		\label{Fig:Plaq}
	\end{subfigure}
	\begin{subfigure}[t]{0.5\linewidth}
		\centring
		\captionsetup{width=.8\linewidth}
		\includesvg[width=1.0\linewidth,pretex=\relscale{0.8}]{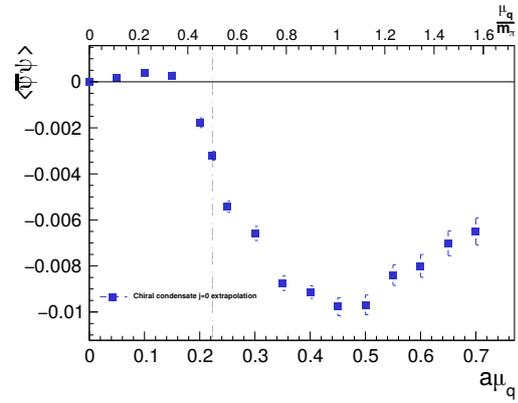}
		\caption{Subtracted and extrapolated chiral condensate.}
		\label{Fig:PBB}
	\end{subfigure}
	\caption{Zero subtracted and diquark extrapolated plaquette sum and chiral condensate.}
	\label{Fig:Sum and chiral}
\end{figure}

\begin{figure}[b]
	\begin{subfigure}[t]{0.5\linewidth}
		\centring
		\captionsetup{width=.8\linewidth}
		\includesvg[width=1.0\linewidth,pretex=\relscale{0.8}]{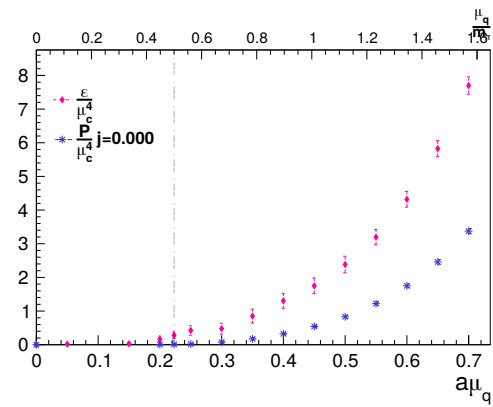}
		\caption{$P$ and $\varepsilon$ normalised by the onset chemical potential $\mu_c=m_\pi$.}
		\label{Fig:Norm_P_e}
	\end{subfigure}
	\begin{subfigure}[t]{0.5\linewidth}
		\centring
		\captionsetup{width=.8\linewidth}
		\includesvg[width=1.0\linewidth,pretex=\relscale{0.8}]{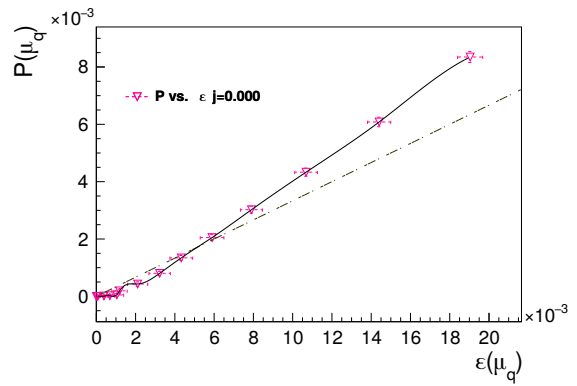}
		\caption{Pressure versus energy density at $aj=0$. Black line is a cubic spline interpolation. Dashed line is the
		conformal limit.}
		\label{Fig:P_v_e}
	\end{subfigure}
	\caption{Pressure and Energy Density}
	\label{Fig:Pressure_eps}
\end{figure}

	The beta functions $\frac{\partial\beta}{\partial a}$ and $\frac{\partial\kappa}{\partial a}$ were evaluated in
	section \ref{Sec:Scale}. Thus all that is left to do is evaluate the plaquette sum and $\langle\bar{\psi}\psi\rangle$
	as seen in figure \ref{Fig:Sum and chiral}. The plaquette sum and $\langle\bar{\psi}\psi\rangle$ depend weakly on the
	diquark source so a linear fit was again used for the zero diquark source extrapolation. We then subtract the $\mu=0$
	values to obtain figure \ref{Fig:Sum and chiral}.
	We can then calculate the full trace anomaly and use equation \eqref{Eq:EnDen} to extract the energy density as seen
	in figure \ref{Fig:P_v_e}.
	\newline
	\subsection{Speed of sound}\label{Sub:C_s}
	Finally, we have all we need to evaluate the speed of sound. We do this by taking the symmetric derivative of
	$P$ with respect to $\varepsilon$. 
\begin{wrapfigure}{O}{0.5\textwidth}
	\centring
	\includesvg[width=1.0\linewidth,pretex=\relscale{0.8}]{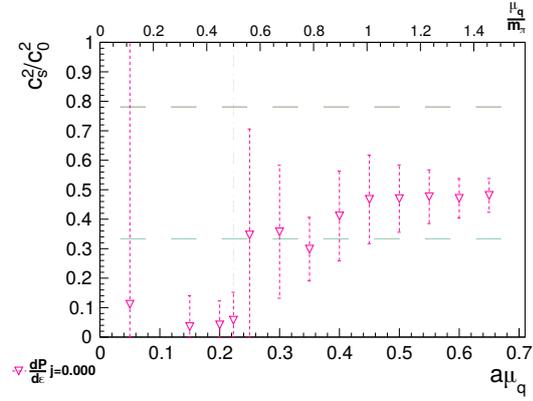}
	\caption{Speed of sound. Upper region denotes area forbidden by relativistic hydrodynamics in
	\cite{Hippert:2024hum}.}
	\label{Fig:C_s}
\end{wrapfigure}
	The results in figure \ref{Fig:C_s} are consistent with other recent \ac{qc2d} works \rjapan{}. There is a sharp
	increase above at the onset chemical potential and $C_s^2$ clearly breaches the conformal limit. This behaviour has also
	been observed in isospin \ac{qcd} simulations. Similarly to \rjapan{} we observe that $C_s^2$ stops behaving like
	the \ac{chiPT} prediction at higher densities and remains well below the bound predicted by relativistic
	hydrodynamics in \cite{Hippert:2024hum}.
\section{Discussions}\label{Sec:Discussions}
	Whilst these early results are promising, there is still more work to do. Measurements have been taken for every
	trajectory, meaning there may be autocorrelations present. The lower diquark source runs were conducted at
	relatively small lattice spatial volume $L_s=16^3$. The error analysis has been crude thusfar. Errors in measured
	observables were bootstrapped rather than jackknifed with multiple elements removed and errors from fits are
	currently read off of the fitting function.

	Moving forward we intend on running all diquark sources at a larger volume with $\sim 1000$ configurations. A run at
	$aj=0.005$ is also underway after upgrades to the \ac{hmc} integrator. Lastly, work is underway to implement a
	Symanzik Improved fermion action to allow further exploration of finer lattice spacings.
\appendix
\section*{Acknowledgements}
\small
This work used the DiRAC Extreme Scaling service (Tursa) at the University of Edinburgh, managed by the Edinburgh
Parallel Computing Centre on behalf of the STFC DiRAC HPC Facility (\url{www.dirac.ac.uk}). The DiRAC service at
Edinburgh was funded by BEIS, UKRI and STFC capital funding and STFC operations grants. DiRAC is part of the UKRI
Digital Research Infrastructure.

Simulations were also performed on the Luxembourg national supercomputer MeluXina. Access was provided through the ICHEC
National Service mechanism. The authors acknowledge the LuxProvide and ICHEC teams for their expert support.
\begin{multicols}{2}
\printunsrtglossary[type=\acronymtype]
\end{multicols}
\bibliographystyle{JHEP}
\bibliography{Proc_Bib}
\end{document}